\documentclass[aps,prl,twocolumn,showpacs]{revtex4}

\usepackage{amssymb}
\usepackage{amsmath}
\usepackage{graphicx}
\usepackage{bm}

\begin{document}

\title{Effects of the $g$-factor in semi-classical kinetic plasma theory}
\author{Gert Brodin}
\affiliation{Department of Physics, Ume{\aa} University, SE--901 87
Ume{\aa}, Sweden}

\author{Mattias Marklund}
\affiliation{Department of Physics, Ume{\aa} University, SE--901 87
Ume{\aa}, Sweden}

\author{Jens Zamanian}
\affiliation{Department of Physics, Ume{\aa} University, SE--901 87
Ume{\aa}, Sweden}

\author{{\AA}sa Ericsson}
\affiliation{Department of Physics, Ume{\aa} University, SE--901 87
Ume{\aa}, Sweden}

\author{Piero L. Mana}
\altaffiliation[Presently at: ]{Perimeter Institute for Theoretical Physics,
31 Caroline st. N., Waterloo, Ontario N2L 2Y5, Canada}
\affiliation{Department of Physics, Ume{\aa} University, SE--901 87
Ume{\aa}, Sweden}

\date{September 13, 2008}

\begin{abstract}
A kinetic theory for spin plasmas is put forward, generalizing those of
previous authors. In the model, the ordinary phase space is extended to
include the spin degrees of freedom. Together with Maxwell's equations, the
system is shown to be energy conserving. Analysing the linear properties, it
is found that new types of wave-particle resonances are possible, that
depend directly on the anomalous magnetic moment of the electron. As a
result new wave modes, not present in the absence of spin, appear. The
implications of our results are discussed.
\end{abstract}

\maketitle

In recent years there has been a rapidly growing interest in the quantum
properties of plasmas \cite
{Haas-2000,Garcia-2005,Shukla-Stenflo-2006,Shukla-Eliasson-2006,Marklund-2007,Brodin-2007,Shukla-2007,Brodin-2008}%
. This has been motivated by applications in, for example, plasmonics \cite
{Atwater-Plasmonics,Marklund-EPL-plasmonics}, quantum wells \cite
{Manfredi-quantum-well} and ultracold plasmas \cite{Ultracold}. Common to
such applications are rather extreme parameters, compared to most laboratory
and space plasmas. In particular the plasma densities are considered to be
very high and/or the temperatures are correspondingly low. For astrophysical
plasmas, it is also known that strong magnetic fields \cite{Astro} may
lead to various quantum effects being important. However, a recent work \cite
{Brodin-2008} show that the spin properties of electrons
%can be important in plasmas even outside such regimes. In particular it was
%found, using a two-fluid model of electrons, that spin effects 
can be
important even in high temperature plasmas of modest density and magnetic
field strength.

In the present Letter we will put forward a more elaborate kinetic
model, where the electrons are described using a distribution function in an
extended phase space, including also variables due to the spin orientation.
This model is an extension of a kinetic model used by Refs. \cite
{Cowley-1986,Kulsrud-1986}, where we here also include the magnetic dipole
force associated with the spin. As a consequence, the magnetic dipole energy
also contributes to the energy conservation law which is derived from
Maxwell's equations combined with the spin-kinetic model. This system is
then used to study linear waves in a homogenous magnetized plasma. The
analysis shows that the inclusion of spin gives rise to new phenomena not present
within the usual Vlasov model. This also holds for the high temperature
regime, where quantum effects are normally suppressed. It turns out that
effects of the anomalous magnetic moment are crucial. In particular, new
wave modes appear with frequencies $\omega \approx (g/2-1)\omega _{c}$,
where $g\simeq 2.002319$ is the electron spin $g$-factor, and $\omega _{c}$ is
the electron cyclotron frequency. Furthermore, new types of wave particle
interaction can take place that involve the electron spin state. We stress
that none of these effects can be seen within quantum fluid models \cite
{Haas-2000,Garcia-2005,Shukla-Stenflo-2006,Shukla-Eliasson-2006,Marklund-2007,Brodin-2007,Shukla-2007,Brodin-2008}.

Here we are interested in effects due to the electron
spin that may survive even when the macroscopic variations
occur on a scale longer than the thermal de Broglie wavelength, which is a
scale normally expected to imply classical behavior \cite{Manfredi-review}. 
%% BEGIN: ADDED IN REVISED VERSION
In particular, we will by spin here mean the semi-classical properties of the
electron due to its magnetic moment, and thus will not take into account, e.g.,  
commutation relations.
%% END: ADDED IN REVISED VERSION
As a starting point, let us consider the evolution equations for momentum
and spin resulting from the Pauli Hamiltonian
\begin{equation}
%H=-\frac{\hbar ^{2}}{2m^{2}}\left( \nabla -\frac{iq\mathbf{A}}{\hbar }%
%\right) ^{2}+\mu _{e}\mathbf{B\cdot }\bm{\sigma }-q\Phi 
H=-(\hbar ^{2}/2m^{2})\left[ \nabla - (iq\mathbf{A}/\hbar )%
\right] ^{2}+\mu _{e}\bm{\sigma }\cdot\mathbf{B}  + q\Phi  ,
\label{Eq-Pauli-Ham}
\end{equation}
where $q = -e = -|e|$ and $m$ are the electron charge and mass respectively, $\mathbf{A}$
and $\Phi $ are the vector and scalar potential, $2\pi \hbar $ is the Planck
constant, $\mu _{e}=-(g/4)e\hbar /m$ is the electron magnetic moment, and $\bm{\sigma }$ is the vector consisting of
the Pauli spin matrices. Using $dF/dt=\partial F/dt+(1/i\hbar )\left[ F,H%
\right] $, where $F$ is any operator, we obtain in the Heisenberg picture 
\begin{subequations}
\begin{eqnarray}
&&\dot{\mathbf{v}}=(q/m)\left( \mathbf{E}+\mathbf{v}\times \mathbf{B}\right)
+(2\mu _{e}/m\hbar )\nabla \left( \mathbf{s\cdot B}\right) ,
\label{Eq-velocity} \\
&&\dot{\mathbf{s}}=(2\mu _{e}/\hbar )\left( \mathbf{s\times B}\right) ,
\label{eq-spin}
\end{eqnarray}
\end{subequations}
where $\mathbf{s}$ is the spin operator, $\mathbf{E}$ and $\mathbf{B}$ are
the electric and magnetic fields,  $\mathbf{v}\equiv \dot{\mathbf{x}}$, and
the over dot denotes the total time derivative. Next, letting the number of
particles with the expectancy value of the velocity between $\mathbf{v}$ and 
$\mathbf{v+}d\mathbf{v}$ and spin vector between $\mathbf{s}$ and $\mathbf{s+%
}d\mathbf{s}$ be given by $dN=fd\mathbf{v}d\mathbf{s}$, we search for an
evolution equation for $f$. Noting that $\nabla _{s}\cdot (\mathbf{
s\times B})=0$, particle conservation implies that the phase space density
is conserved along fluid elements propagating in the extended phase space, i.e., 
$\dot{f}(\mathbf{r},\mathbf{v},\mathbf{s}%
,t)=\partial _{t}f+\dot{\mathbf{r}}\cdot \nabla f+\dot{\mathbf{v}}\cdot
\nabla _{v}f+\dot{\mathbf{s}}\cdot \nabla _{s}f=0$. Thus, with Eqs.\ (2),
\begin{eqnarray}
&&\!\!\!\!\!\!\!\!\partial _{t}f+\mathbf{v\cdot }\nabla f+\left[ \frac{q}{%
m}\left( \mathbf{E}+\mathbf{v}\times \mathbf{B}\right) \mathbf{+}\frac{2\mu
_{e}}{m\hbar }\nabla \left( \mathbf{s\cdot B}\right) \right] \cdot \nabla
_{v}f  \nonumber \\
&&\qquad +\frac{2\mu _{e}}{\hbar }\left( \mathbf{s\times B}\right) \cdot \nabla
_{s}f = 0 . 
%&&\!\!\!
%\partial _{t}f+\mathbf{v\cdot }\nabla f+\left[ (q/m)\left( \mathbf{E}+\mathbf{v}\times \mathbf{B}\right) + (2\mu
%_{e}/m\hbar )\nabla \left( \mathbf{s\cdot B}\right) \right] \cdot \nabla
%_{v}f  \nonumber \\
%&&+ (2\mu _{e}/\hbar )\left( \mathbf{s\times B}\right) \cdot \nabla
%_{s}f  = 0.
\label{Eq-f-final}
\end{eqnarray}
For the concept of a distribution function in phase space to be well
defined, the delocalization of the wave function of the particles cannot be
too large. To establish the limit of validity, we therefore would like to
connect (\ref{Eq-f-final}) directly to the Pauli equation. Studying the
one-particle equation $i\hbar \partial _{t}\Psi _{\alpha }=H\Psi _{\alpha }$%
, where $\Psi _{\alpha }$ is the electron wave function and the subscript $\alpha $ is a particle index, the Pauli
equation can be transformed into fluid like variables by the transformation $%
\Psi _{\alpha }=\sqrt{n_{\alpha }}\,\exp(iS_{\alpha}/\hbar )\varphi _{\alpha}$,
where $\varphi _{\alpha}$ is a unit normalized two-spinor. Defining $\mathbf{v}%
_{\alpha }=(1/m)(\nabla S_{\alpha}-i\hbar \varphi _{\alpha }^{\dagger }\nabla
\varphi _{\alpha })-(q/m)\mathbf{A}$ and $\mathbf{s}_{\alpha }=(\hbar
/2)\varphi _{\alpha }^{\dagger }\bm{\sigma }\varphi _{\alpha }$, the Pauli
equation can be rewritten as evolution equations for $n_{\alpha }$, $\mathbf{v}%
_{\alpha }$ and $\mathbf{s}_{\alpha }$ \cite{Marklund-2007}, which have a
fluid like form, although we are still dealing with single particle
equations. These equations
resemble Eqs. (\ref{Eq-velocity}) and (\ref{eq-spin}) above, but contain
several other terms due to the difference of a two-spinor from a classical
particle with a magnetic moment. In particular, a force-like term that
arises in this way, even in the absence of spin, is the so called Bohm
potential which results in a force $\mathbf{F}_{B}=(\hbar ^{2}/2m)\nabla
[ (\nabla ^{2}\sqrt{n_{\alpha }})/\sqrt{n_{\alpha }} ] $. Assuming
a characteristic (thermal) energy of the particles, the effect of the Bohm
potential is seen to be small provided the gradient scale length
is longer than the thermal de Broglie wavelength \cite{Manfredi-review}.
Similarly, quantum corrections to the Vlasov equation due to wave function dispersion 
\cite{Manfredi-review}, has been shown to be small
whenever spatial gradients are long compared to this quantum scale.
Furthermore, the spin properties induces higher order force terms $%
F_{j}=(1/m)\partial ^{i}\left[ (\partial _{j}s_{k})(\partial _{i}s^{k})%
\right] $ \cite{spin-review} that are small (compared to the
convective derivative), provided the gradient scale length is longer than
the thermal de Broglie wavelength. Finally, higher order terms in the spin
evolution equation \cite{spin-review} resulting from the Pauli equation, can
be neglected when the same condition is fulfilled. The kinetic equation put
forward here agrees with the one used by Refs. \cite
{Cowley-1986,Kulsrud-1986}, except that their equation did not contain the
magnetic dipole force. This term is often small as compared to the Lorentz
force, but we will show that it can still be important.

A complete model is formed by combining Eq. (\ref{Eq-f-final}) with Maxwells
equations, where the current density is 
\begin{equation}
\mathbf{j}=\mathbf{j}_{\mathrm{free}}+\nabla \times \mathbf{M}=\sum_{i}\left[
q_{i}\!\!\int \mathbf{v}f_{i}d\Omega +\frac{2\mu_{i}}{\hslash }\nabla \times
\!\!\int \mathbf{s}f_{i}d\Omega \right] .  \label{Eq-current}
\end{equation}
The sum is over particle species $i$ with charge $q_i$, and the last term is the
magnetization current due to the spin. Normally the spin contribution from
the ions can be neglected compared to that of the electrons, due to their
smaller magnetic moment. In what follows, we will therefore only include the
electron physics, and drop the sum over species. The integration $d\Omega $
is made over three velocity variables and two spin degrees of freedom. Since
the spin vector is constructed from the expectancy value of the spin
operator, it has a fixed length $\left| \mathbf{s}\right| =\hbar /2$.
The spin orientation will be described using spherical
coordinates. %$\varphi _{s}$ and $\theta _{s}$ (where we use index $%
%s $ in order not to confuse them with angles in velocity space). 
From this model it is straightforward to show that the energy
conservation law 
$%\begin{equation}
\partial _{t}W+\nabla \cdot \mathbf{P=0}  \label{eq-e-cons}
$ %\end{equation}
is fulfilled, with the energy density $W$ and energy flux $\mathbf{P}$ given
by 
$%\begin{equation}
 W = \left( \varepsilon _{0}E^{2}+ \mu_0^{-1}B^{2}\right)/2 -%
\mathbf{B}\cdot \mathbf{M}+\int (mv^{2}/2)fd\Omega   \label{eq-edens}
$ %\end{equation}
and 
$%\begin{equation}
\mathbf{P} = \mathbf{E\times }\left( \mu_0^{-1}\mathbf{B} - \mathbf{M}%
\right)\!\! + \!\!\int \left[ (mv^{2}/2)\mathbf{v}    
%\nonumber \\ && 
- (2\mu _{e}/{\hbar}) \mathbf{v}(\mathbf{B}\cdot \mathbf{s)}) \right]
fd\Omega   \label{eq-eflux}
$, %\end{equation}
respectively. The last term in the energy flux expression represents the convection of
magnetic dipole energy.

Next we will use Eq. (\ref{Eq-f-final}) to study linear waves in a
magnetized plasma. Dividing the variables as $f=f_{0}(\mathbf{v},\mathbf{s}%
)+f_{1}(\mathbf{r},t,\mathbf{v},\mathbf{s})$, and $\mathbf{B}=\mathbf{B}_{0}+%
\mathbf{B}_{1}(\mathbf{r},t)$, the linearized Vlasov equation can be written 
\begin{eqnarray}
&&\!\!\!\!\!\!\!\!\left[ \partial _{t}+\mathbf{v\cdot }\nabla +\frac{q}{m}%
\left( \mathbf{v}\times \mathbf{B}_{0}\right) \cdot \nabla _{v}+\frac{2\mu
_{e}}{\hbar }\left( \mathbf{s\times B}_{0}\right) \cdot \nabla _{s}\right]
f_{1}=  
\nonumber \\ &&\!\!\!\!
- \frac{q}{m}\left( \mathbf{E+v}\times \mathbf{B}_{1}\right) \cdot \nabla _{v}  f_0
\nonumber \\ &&
- \frac{2\mu _{e}}{\hbar }\left[\frac{\nabla \left( \mathbf{s\cdot B}_{1}\right)}{m}\cdot \nabla _{v}
+ \left( \mathbf{s\times B}_{1}\right) \cdot \nabla
_{s}\right]f_{0} 
.  \label{eq:f-linear-1}
\end{eqnarray}
Before proceeding further, we will for definiteness specify the unperturbed
equilibrium distribution $f_{0}$. In thermodynamic equilibrium and for a
large chemical potential (which applies for $n\lambda _{dB}^3 \gg 1$, where $%
\lambda _{dB}$ is the thermal de Broglie wavelength, and $n$ is the electron
number density), the Fermi-Dirac equilibrium distribution reduces to
\begin{equation}
\! f_{0} =\frac{n_{0}}{4\pi }\left( \frac{m}{2\pi k_{B}T}\right)
^{3/2}\!\!\!\!\!\!\exp \left[-\frac{(mv^{2}/2+2\mu _{e}\mathbf{s\cdot B}_{0}/\hbar )}{k_{B}T}\right] .
\label{eq-f0}
\end{equation}
This results in a zero order magnetization $\mathbf{M}_{0}=n_{0}\mu
_{e}\eta \left( {\mu _{e}B_{0}}/{k_{B}T}\right) $, where $\eta $ is the
Langevin function, $T$ is the temperature, $n_{0}$ the equilibrium density
and $k_{B}$ is the Boltzmann constant. %For most laboratory plasma 
%applications, the argument of the Langevin function is small, which
%correspond to a small dipole energy (compared to the thermal energy), and
%consequently there is only a slight overweight of particles in the lowest
%energy spin state. For this case we may use the first order Taylor expansion 
%$\eta \simeq \mu _{e}B_{0}/k_{B}T$. 
Next, letting $\mathbf{B}_{0}=B_{0}\widehat{\mathbf{z}}$, introducing
cylindrical coordinates  $(v_{\bot }, \varphi _{v}, 
v_{z})$ in velocity space and spherical coordinates  $(\varphi _{s}, \theta _{s}
)$ in spin space, noting that the term $(\mathbf{v}\times \mathbf{B}_{1})\cdot \nabla
_{v}f_{0}$ in (\ref{eq:f-linear-1}) can be dropped when the unperturbed
distribution function is Maxwellian, and Fourier analyzing, Eq. (\ref
{eq:f-linear-1}) is written 
\begin{eqnarray}
&&\!\!\!\!\!\!\!\!\!\!\!\!\!\!\!\left[ i\left( \omega -\mathbf{k\cdot v}%
\right) \mathbf{+}\omega _{c}\frac{\partial }{\partial \varphi _{v}}+\omega
_{cg}\frac{\partial }{\partial \varphi _{s}}\right] \widetilde{f}_{1}= 
\nonumber \\
&&\!\!\!\!\!\!\!\!\!\!\!\!\!\!\!\left[ \frac{q}{m}\widetilde{\mathbf{E}}%
\mathbf{+}\frac{2\mu _{e}}{m\hbar }\nabla (\mathbf{s\cdot }\widetilde{%
\mathbf{B}}_{1})\right] \cdot \nabla _{v}f_{0}+\frac{2\mu _{e}}{\hbar }(%
\mathbf{s\times }\widetilde{\mathbf{B}}_{1})\cdot \nabla _{s}f_{0}
\label{eq-flinear2}
\end{eqnarray}
where we have introduced  the frequencies $\omega _{c}=qB_{0}/m$ and $\omega _{cg}=2\mu _{e}B_{0}/\hbar $. Note that $\omega _{c}<0$ and that $\omega
_{cg}=(g/2)\omega _{c}$. Eq. (\ref{eq-flinear2}) can be solved by an
expansion of $\widetilde{f}_{1}$ in the eigenfunctions  
$
\psi _{a}(\varphi _{v},v_{\bot })= (2\pi)^{-1/2}\exp [-i(a\varphi
_{v}-k_{\bot }v_{\bot }\sin \varphi _{v}/\omega _{c})] % \label{eq-eigenfunc}
$. 
Thus, we let 
\begin{equation}
\widetilde{f}_{1}=\sum_{a,b}g_{ab}(v_{\bot },v_{z},\theta _{s})\psi
_{a}(\varphi _{v},v_{\bot })\exp (-ib\varphi _{s})  \label{eq-expansion}
\end{equation}
where $a=0,\pm 1,\pm 2,...$ and $b=-1,0,1$. Using the orthogonality properties $\int_{0}^{2\pi }\psi
_{a}\psi _{b}^{\ast }d\varphi _{v}=\delta _{ab}$ we find 
\begin{equation}
i\left( \omega -k_{z}v_{z} - a\omega_{c} - b\omega _{cg}\right)
g_{ab}=I_{ab}(v_{\bot },v_{z},\theta _{s})  \label{eq_res-factor}
\end{equation}
with 
\begin{eqnarray}
I_{ab} &=&\!\!\!\!\!\!\int_{0}^{2\pi }\int_{0}^{2\pi }\left\{ \left[ \frac{q%
}{m}\widetilde{\mathbf{E}} + \frac{2\mu _{e}}{m\hbar }\nabla (\mathbf{%
s\cdot }\widetilde{\mathbf{B}}_{1})\right] \cdot \nabla _{v}f_{0}\right.  
\nonumber \\
&&\!\!\!\!\!\!\!\!\!\!\!\!\left. +\frac{2\mu _{e}}{\hbar }(\mathbf{s\times }%
\widetilde{\mathbf{B}}_{1})\cdot \nabla _{s}f_{0}\right\} \psi _{a}^{\ast
}\exp (ib\varphi _{s})d\varphi _{v}d\varphi _{s}.  \label{eq-I-coeff}
\end{eqnarray}
A useful relation when trying to write results in a more explicit
form is the Bessel-expansion 
\begin{equation}
\psi _{a}(\varphi _{v},v_{\bot })=\frac{1}{\sqrt{2\pi }}\sum_{b}J_{b}\left(\! \frac{%
k_{\bot }v_{\bot }}{\omega _{c}}\!\right) \exp [i(b-a)\varphi _{v}].
\label{eq:Bessel-exp}
\end{equation}
Here it is seen that the results are much simplified in the limit where $%
k_{\bot }v_{th}/\omega _{c}$ is small (where we can estimate $v_{\bot }$
with the thermal velocity $v_{th}$), but as is well-known (see e.g. Ref. 
\cite{Swanson}), in general the conductivity tensor components turns into
sums over Bessel functions . The conductivity tensor $\sigma^{ij}$, defined by $%
j^{i}=\sigma ^{ij}E_{j}$, is found from Eq. (\ref{Eq-current}) by expressing
the magnetic field in terms of $\widetilde{\mathbf{E}}$ and then solving for 
$\widetilde{f}_{1}$ in terms of $\widetilde{\mathbf{E}}$ using the
eigenfunctions as outlined above. It is illustrative to divide the
conductivity tensor in two contributions $\sigma ^{ij}=\sigma _{\mathrm{free}%
}^{ij}+\sigma _{\mathrm{magn}}^{ij}$ due to the free current and the
magnetization current. Furthermore $\sigma _{\mathrm{free}}^{ij}$ could be
divided further into contributions due to the Lorentz force and
contributions in $I_{ab}$ that contains spin which give the classical and
spin parts of $\sigma _{\mathrm{free}}^{ij}$. However, we will not present
the general expressions for $\sigma ^{ij}$ here, as these results are
complicated and needs extensive analysis for a useful interpretation.
Instead we focus on special cases in order to point out two of the main new
features resulting from the spin.

Firstly, the factor $\left( \omega -k_{z}v_{z} - a\omega _{c} - b\omega
_{cg}\right) $ in (\ref{eq_res-factor}) reveals that the standard
wave-particle resonances are extended to involve the spin-forces. The
spectrum of resonances is thereby much extended. In particular the
combination $a=1,b=-1$ give a resonant velocity at $v_{z}=(\omega -\Delta
\omega _{c})/k_{z}$, where we have introduced $\Delta \omega
_{c}=(g/2-1)eB_{0}/m$. The physics of this resonance is slightly more
complicated than the well-known resonances with $b=0$. In particular, it
does not occur for strictly parallel propagation to $\mathbf{B}_{0}$, since
the coefficient $I_{1-1}$ in (\ref{eq-I-coeff}) becomes zero in that limit.
Furthermore, inclusion of the spatial variations of the wave perpendicular
to the magnetic field must include a finite argument of the Bessel functions
in (\ref{eq:Bessel-exp}). This means that the spin-resonances depend on
finite Larmor radius effects. Moreover, integration over $\varphi _{s}$
kills any contribution from this resonance in the free current density, and
thus only the magnetization current survives. Still it is clear that such
resonances can be the dominating wave-particle damping mechanism, in case
the wave frequency is of the order of a few thousands of the cyclotron
frequency.

The second point to be pursued in more detail is that the spin
gives rise to new wave modes not present in the standard Vlasov
picture. Firstly, we assume that the ions are immobile, constituting a
neutralizing background. Next, we consider the limit of exactly
perpendicular propagation, and thus let $\mathbf{k}=k_{\bot }\widehat{%
\mathbf{x}}$. In this case modes with the approximate polarization $%
\widetilde{\mathbf{E}}=\widetilde{E}_{z}\widehat{\mathbf{z}}$ and $%
\widetilde{\mathbf{B}}=\widetilde{B}_{y}\widehat{\mathbf{y}}=(k_{\bot }%
\widetilde{E}_{z}/\omega )\widehat{\mathbf{y}}$ are possible, provided $%
\sigma ^{xy}$ and $\sigma ^{xz}$ are sufficiently small, which can be
verified \textit{a posteriori} (see, e.g., Ref.\ \cite{Swanson}). 
From Eqs. (\ref{eq-expansion}), (\ref{eq_res-factor}) and (\ref
{eq-I-coeff}) the distribution function is expressed in terms of the
electric field, that together with (\ref{Eq-current}) determines the
conductivity tensor components. For the given polarization, only the $\sigma
^{zz}$ component is needed. Furthermore, for $\omega \ll \left| \omega
_{c}\right| $, terms with coefficients $a \neq b$ in (\ref{eq-expansion})
will be smaller than those with $a = b$, and thus only the latter are kept.
Carrying out the $\varphi _{s}$ and $\varphi _{v}$ integrals [using (\ref
{eq:Bessel-exp})], the $z$-component of Ampere's law
gives
\begin{eqnarray}
&& 
\omega ^{2} =  k^{2}c^{2}+\omega _{p}^{2}\int \Big\{J_{0}^{2}\left( {%
k_{\bot }v_{\bot }}/{\omega _{c}}\right)  
\nonumber \\ &&
+\frac{k^{2}\hbar ^{2}\Delta \omega _{c}\sin ^{2}\theta
_{s}}{4m(\omega -\Delta \omega _{c})k_{B}T}\left[ J_{1}^{2}\left( {%
k_{\bot }v_{\bot }}/{\omega _{c}}\right) \right] \Big\}f_{0}\,d\Omega ,
\label{Eq-DR}
\end{eqnarray}
where $\omega _{p}=(n_{0}e^{2}/\epsilon _{0}m)^{1/2}$ is the plasma
frequency. If we drop the term proportional to $J_{1}^{2}$ due to the spin
in (\ref{Eq-DR}), we have the usual ordinary mode, where higher terms in the
sum over Bessel functions have been dropped, due to the condition $\omega
\ll \left| \omega _{c}\right| $. Typically the second term due to the spin
is a small correction when the wavelength is longer than the thermal de
Broglie wavelength. However, it can clearly be dominant for frequencies
close to the resonance $\omega \approx \Delta \omega _{c}$. In Fig 1
numerical solutions of the dispersion relation is plotted for different
parameters $\zeta $ and $\eta $ close to this resonance, where $\zeta =\hbar
^{2}\omega _{c}^{2}/m^{2}v_{t}^{4}$, $\eta =\omega _{c}^{2}c^{2}/\omega
_{p}^{2}v_{t}^{2}$ and $v_{t}=\left( k_{B}T/m\right) ^{1/2}$. \ To some
extent these \ new solutions resembles the well-known Bernstein modes (see
e.g. Ref. \cite{Swanson}). However, unless the ratio $\zeta /\eta $ is
larger than unity (corresponding to a high density low temperature plasma),
the deviation from the resonant frequency will be small. For an unperturbed
distribution different from the thermodynamic equilibrium expression the
shift from the precise resonance can be much enhanced, since the two last
terms in (\ref{eq-I-coeff}) contributing to $I_{1-1}$ almost cancel (the
terms match as $g/2-1$) for $f_{0}$ given by (\ref{eq-f0}) as considered
here, but not for a general equilibrium distribution. Furthermore, it should
be noted that this approximate cancellation in (\ref
{eq-I-coeff}) does not occur for general angles of
propagation. For arbitrary angles of propagation, the denominator $\omega
-\Delta \omega _{c}$ of our spin term is replaced by $\omega
-k_{z}v_{z}-\Delta \omega _{c}$. It is interesting to note that the sign of
this contribution is determined by the last term in (\ref{eq-I-coeff}). In
particular, in case the unperturbed distribution $f_{0}$ has a larger
fraction of particles in the higher energy spin state, $\partial
f_{0}/\partial s_{z}$ changes sign, which should consequently change the
sign of the imaginary part of the dispersion relation, leading to
instability rather than wave-particle-spin damping. However, more analysis
is needed to definitely confirm this conjecture.

%%%%%% FIGS %%%%%%
\begin{figure}[tbh]
\includegraphics[width=.8\columnwidth]{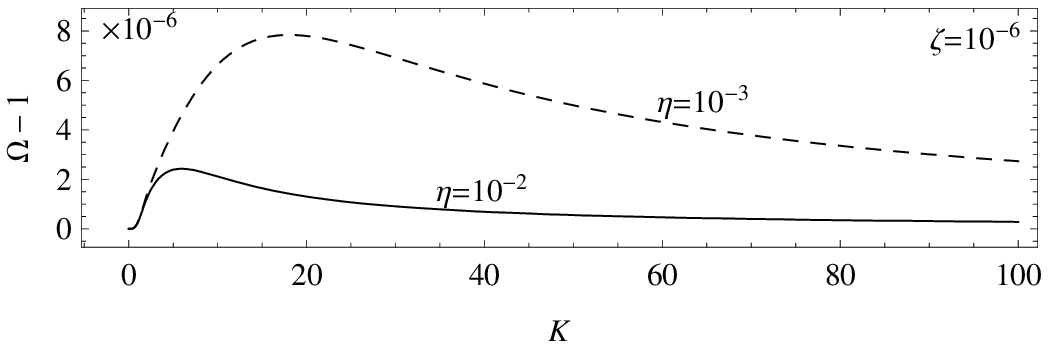} %
\includegraphics[width=.8\columnwidth]{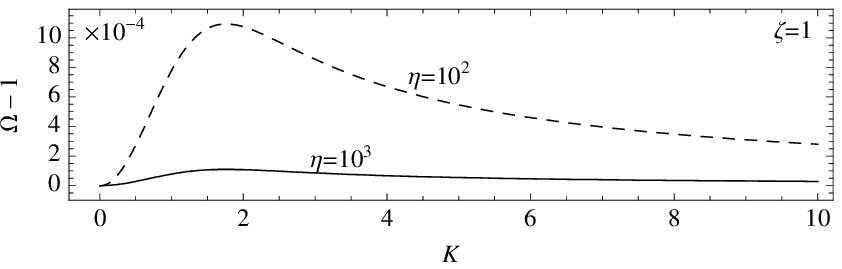}
\caption{
%The normalized frequency $\Omega = \protect\omega/\Delta\protect%
%\omega_c$ plotted around $\Delta\protect\omega_c$ as a function of the
%normalized wavenumber $K =kv_t/\protect\omega_c$ for different parameter
%values of the parameters $\protect\zeta=\hbar ^{2}\protect\omega
%_{c}^{2}/m^{2}v_{t}^{4}$ and $\protect\eta=\protect\omega _{c}^{2}c^{2}/%
%\protect\omega _{p}^{2}v_{t}^{2}$. The upper panel represents possible
%laboratory values, while the lower are representative for extreme
%astrophysical environments.
The normalized frequency $\Omega =\protect\omega /\Delta \protect%
\omega _{c}$ plotted around $\Delta \protect\omega _{c}$ as a function of
the normalized wavenumber $K=kv_{t}/\protect\omega _{c}$ for different
parameter values of the parameters $\protect\zeta =\hbar ^{2}\protect\omega
_{c}^{2}/m^{2}v_{t}^{4}$ and $\protect\eta =\protect\omega _{c}^{2}c^{2}/%
\protect\omega _{p}^{2}v_{t}^{2}$. The upper panel represents possible
laboratory values (for example, $B_{0}=6.0\mathrm{T}$, $T=10^{4}\mathrm{K,}$ and $n=10^{22}%
\mathrm{cm}^{-3}$ correspond to $\zeta =10^{-6}$ and $\eta =10^{-2}$,
relevant for laser-plasma interaction experiments), while the lower are representative for extreme
astrophysical environments (for example, $%
B_{0}=6.0\times 10^{6}\mathrm{T}$, $T=10^{10}\mathrm{K,}$ and $n=3\times
10^{21}\mathrm{cm}^{-3}$ correspond to $\zeta =1$ and $\eta =10^{-3}$,
relevant for a thick hot accretion disk surrounding a pulsar).}
\end{figure}
%%%%%%%%%%%%%%%

In the present Letter we have put forward a Vlasov-type equation, in a
phase space extended to include the spin degrees of freedom. This equation
extends a previous equation due to Refs \cite{Cowley-1986,Kulsrud-1986}, by
including the magnetic dipole force associated with the spin, and can be
viewed as a semi-classical limit of the Pauli equation. Including the
spin-magnetization current in\ Maxwells equations, the resulting system is
shown to be energy conserving. To gain some understanding of the model, we
have outlined the theory for linear wave propagation in magnetized plasmas
and demonstrated the appearance of new wave modes due to the spin.
These wave modes depend on resonances associated with both the orbital and
spin gyration, and is much different from the well-known spin waves in
ferromagnetic materials \cite{Spin-waves}.

It is of interest to discuss the connection between our spin-kinetic model and the
ordinary Vlasov equation. Relating $\mathbf{E}$ and $\mathbf{B}$ through
Faraday's law, we see that the relative strength of the spin force is $%
k^{2}\hbar /m\omega $. This parameter is often small, which explains why
the classical Vlasov equation in many cases is a very good approximation. On
the other hand, using this parameter it is easy to underestimate the
significance of the spin force. Firstly, wave-particle resonances and/or new
wave modes associated with resonances, as the one described above, may
enhance the significance of spin effects. Secondly, when the electric field
is perpendicular to $\mathbf{B}_{0}$ and the wave frequencies are low, the $%
\mathbf{E\times B}$-motion of electrons and ions give a zero current to
leading order in $\omega /\omega _{c}$, whereas similar cancellations does
not occur for the magnetic dipole force. Furthermore, even if the electric
field is perpendicular to $\mathbf{B}_{0}$, the dipole force may have a
parallel component leading to a large current in this direction.

The exploration of Eq.\ (\ref{Eq-f-final}) is a very rich problem. Besides the
more obvious generalization to a complete linear theory, including arbitrary
angles of propagation, there is a large set of nonlinear problems that
should be studied. The results by Ref. \cite{Brodin-2008} indicate that the
significance of the spin force can be increased in the nonlinear regime as
compared to the linear one, when the spin dependence of the ponderomotive
force can separate particles of different spin. Finally, it should be
stressed that the current model is intended for the ''weak quantum'' regime
where the characteristic length scale is longer than the thermal de Broglie
wavelength, and the Zeeman energy is smaller than the thermal energy. A very
interesting problem is generalizations to the regime where the Zeeman energy
is comparable or larger than the thermal energy. For such parameters, found in astrophysical settings, effects such as Landau quantization 
\cite{Melrose2006} will enter the picture.

\acknowledgments
This research was supported by the Kempe Foundations through the contract SMK-2647, by the European Research Council under Contract No.\ 204059-QPQV, and the Swedish Research Council under  
Contracts No.\ 2005-4967 and No.\ 2007-4422 .

\end{document}